\documentclass[twocolumn,trackchanges]{aastex701}
\usepackage{graphicx}
\usepackage{subfigure}
\usepackage{paralist}
\usepackage{amssymb}
\usepackage{bm}
\usepackage{bbding}
\usepackage[fleqn]{amsmath}
\usepackage{multirow}
\hypersetup{linkcolor=blue,filecolor=blue,urlcolor=blue}

\begin{document}

\title{Eccentricity and Inclination Excitation of Stellar Orbits around merging Black-Hole Binaries}

\correspondingauthor{Hanlun Lei}

\author{Hao Gao} 
\affiliation{School of Astronomy and Space Science, Nanjing University, Nanjing 210023, China}
\affiliation{Key Laboratory of Modern Astronomy and Astrophysics in Ministry of Education, Nanjing University, Nanjing 210023, China}
\email{1528339424@qq.com}

\author{Hanlun Lei}
\affiliation{School of Astronomy and Space Science, Nanjing University, Nanjing 210023, China}
\affiliation{Key Laboratory of Modern Astronomy and Astrophysics in Ministry of Education, Nanjing University, Nanjing 210023, China}
\email[show]{leihl@nju.edu.cn}

%

\begin{abstract}
Identifying the formation channels of merging compact-object binaries remains a fundamental challenge in gravitational-wave (GW) astrophysics. Here, we formulate a generalized framework incorporating both third-body perturbations and post-Newtonian effects to investigate the resonance-driven excitation of eccentricity and/or inclination for stellar orbits around merging black-hole binaries (BHBs). We show that at the quadrupole level, inclination excitation is driven by adiabatic capture into a quadrupole-order resonance, with subsequent resonance escape governed by the conservation of phase-space area. Furthermore, in the coplanar octupole regime, an apsidal precession resonance triggers robust eccentricity growth, where the capture probability depends on the initial conditions. For spatial octupole configurations, we reveal a distinct two-stage evolutionary pathway: the quadrupole-order resonance first excites the orbital inclination, followed by subsequent eccentricity growth driven by the inverse von Zeipel--Lidov--Kozai (ZLK) resonance. Our results indicate that these resonance-driven dynamics could imprint observable kinematic signatures on surrounding stellar populations, thereby offering a novel indirect probe for merging BHBs.
\end{abstract}



\section{ Introduction} 

The origin of merging compact-object binaries remains a central problem in GW astrophysics. Since the first detection of GWs from binary black hole mergers by the LIGO--Virgo--KAGRA collaboration \citep{abbott2016observation}, extensive efforts have been devoted to identifying their formation channels and dynamical pathways. Proposed scenarios include isolated binary evolution in galactic fields \citep{lipunov1997formation,podsiadlowski2003formation,dominik2012double,dominik2013double,dominik2015double,belczynski2016first}, dynamical assembly in dense stellar systems \citep{o2006binary,rodriguez2015binary,antonini2016merging,rodriguez2016binary,antonini2019black}, and multi-body interactions in hierarchical systems \citep{dotti2012massive,antonini2017binary,fragione2019black,liu2019enhanced,martinez2020black}. Because these channels can produce distinct orbital signatures, orbital dynamics serves as a powerful diagnostic tool for probing the origins of GW sources.

The rich dynamics of hierarchical triple systems provide a natural and efficient mechanism for driving orbital excitation through long-term secular interactions \citep{ford2000secular,naoz2013secular}. A substantial fraction of compact binaries is thought to reside in triple configurations, particularly within dense stellar environments \citep{sana2012binary}. Recent observations identifying BHBs in hierarchical triples lend compelling empirical support to such configurations \citep{burdge2024black}. Furthermore, population synthesis and dynamical studies show that triple‑induced secular evolution offers a robust framework to explain the formation of wide and eccentric BHBs \citep{generozov2024triple}; these studies also indicate that a significant proportion of observed compact binaries might originate from triples with a prior merger history \citep{shariat2025once}. Moreover, recent numerical simulations indicate that three‑body interactions in dense environments or compact triples can efficiently excite high eccentricities and accelerate mergers \citep{trani2024isles,vigna2025prompt,barber2025formation}. At the quadrupole level, the ZLK mechanism can induce large-amplitude coupled oscillations in eccentricity and inclination, provided the mutual inclination exceeds a critical threshold \citep{von1910application,lidov1962evolution,kozai1962secular,ito2019lidov}. This mechanism has been extensively studied and applied to a wide range of astrophysical systems, including exoplanets, stellar binaries, and compact-object mergers \citep{fabrycky2007shrinking, antonini2012secular,petrovich2015steady,naoz2016eccentric}.

Beyond the quadrupole approximation, higher-order secular effects significantly enrich the dynamics. In particular, octupole-level perturbations break the integrability of the system and introduce qualitatively different behaviors, such as extreme eccentricity excitation, orbital flips, and chaotic evolution \citep{naoz2013secular,lithwick2011theory,li2014chaos,katz2011long,lei2022systematic,lei2022dynamical}. These effects are especially important in systems with unequal masses and finite outer binary eccentricities. Coplanar octupole-order dynamics has also been explored in planetary systems, where secular interactions can drive high-eccentricity migration \citep{petrovich2015steady,xue2017possible}, while the secular dynamics of outer test particles in hierarchical triples has also been explored  \citep{bhaskar2021mildly}.

In addition to inclination-driven excitation, recent studies have highlighted alternative resonance mechanisms that operate even in nearly coplanar configurations. A notable example is the apsidal precession resonance, which occurs when the apsidal precession rates of the inner and outer orbits become commensurate, leading to eccentricity excitation \citep{liu2015merging,liu2020merging,liu2022uncovering,kuntz2022precession,liu2024extreme,farhat2025capture}. This mechanism provides a complementary channel to the classical ZLK effect and opens a new window for probing compact binaries in low-inclination systems.

When dissipation is included, the secular dynamics becomes intrinsically time-dependent. In this regime, the system evolves adiabatically and its evolution can be described in terms of resonance capture, adiabatic invariance, and separatrix crossing \citep{peters1964gravitational,henrard1982capture,neishtadt1984separation}. Related resonance-capture processes have also been studied in slowly evolving compact object systems \citep{munoz2022eccentric,bhaskar2022black}. In this regard, it is demonstrated that apsidal precession resonances can efficiently pump eccentricity of the outer binaries in coplanar hierarchical systems under GW and/or tidal dissipation \citep{liu2024extreme, farhat2025capture}. However, the dynamical behavior in a general configuration with dissipation, particularly those with multiple degrees of freedom, remains unexplored.

In this work, we extend the coplanar scenario to a general framework. In particular, we systematically investigate three configurations: the quadrupole, the coplanar octupole\footnote{The case of coplanar triple was explored by \citet{liu2024extreme}. In the present study, we include it for the purpose of completeness.}, and the spatial octupole cases. The eccentricity and inclination excitation of stellar orbits around merging BHBs is investigated using both numerical and analytical methods. In particular, the evolution of orbital elements induced by the quadrupole- and/or octupole-order resonances may serve as an indirect probe of embedded BHBs and GW progenitors. 

The remaining part of this study is organized as follows. In Section \ref{sec:2}, we briefly introduce the dynamical model with GW dissipation. Then, resonance-driven dynamical features are explored in the quadrupole (Section \ref{sec:3}), coplanar octupole (Section \ref{sec:4}), and spatial octupole (Section \ref{sec:5}) configurations. At last, we summarize our main findings in Section \ref{sec:6}. 

\section{Dynamical model} \label{sec:2}

In this work, we investigate the secular dynamics of an exterior particle within a hierarchical three-body system. This restricted model serves as a good approximation for describing stellar orbits around BHBs \citep{liu2024extreme}. The inner system consists of two black holes with masses $m_1$ and $m_2$, forming a compact relativistic binary. The orbital evolution of this inner BHB is governed by both conservative and dissipative relativistic corrections. Specifically, the inner BHB undergoes apsidal precession at the first post-Newtonian (1PN) order, and GW emission (at the 2.5PN order) drives BHB's orbital decay \citep{peters1964gravitational,blanchet2014gravitational,maggiore2008gravitational}. 

For convenience, we adopt the barycenter invariant-plane reference frame, in which the motion of celestial bodies is described by the classical orbital elements $(a,e,i,\Omega,\omega,M)$. The longitude of pericenter is defined as $\varpi = \Omega + \omega$. Unless otherwise stated, the elements with subscript $1$ are for the inner BHB, and the ones with subscript $2$ are for the outer particle. In hierarchical configurations, the semimajor axis ratio $\alpha=a_1/a_2$ is a small parameter such that the Hamiltonian function can be expanded as a power series in $\alpha$. In practice, we truncate the Hamiltonian at the octupole order of $\alpha=a_1/a_2$. 

To study the long-term evolution of stellar orbits, it is usual to average the Hamiltonian over the orbital periods of the inner and outer binaries \citep{ford2000secular,naoz2013secular,naoz2016eccentric}. The resulting Hamiltonian is given by \citep{naoz2017eccentric, vinson2018secular, de2019inverse, lei2024dynamical}
\begin{equation}\label{Eq2}
{\cal H}_{\rm sec}  = -{\cal C}_0(F_{\rm{quad}}+ {\varepsilon} F_{\rm{oct}}),
\end{equation}
where ${\cal C}_0$ and $\varepsilon$ are defined as
\begin{equation*}
{\cal C}_0=\frac{{\cal G}m_1m_2}{16(m_1+m_2)}\frac{a_1^2}{a_2^3},\; \varepsilon = \frac{m_1-m_2}{m_1+m_2}\frac{a_1}{a_2},
\end{equation*}
with $\cal G$ as the universal gravitational constant. The explicit forms of the quadrupole and octupole terms ($F_{\rm{quad}}$ and $F_{\rm{oct}}$) are provided in Appendix \ref{sec:app}. 

Accounting for the leading-order relativistic effects, the pericenter precession of the inner BHB is governed by
\begin{equation}\label{Eq5}
{\dot \varpi _1} = \frac{{3n_1}}{\eta_1^2}\frac{{{\cal G}m_{12}}}{{a_1{c^2}}},
\end{equation}
where $c$ is the speed of light, $n_1$ is the mean motion of the inner BHB, and $\eta_1 = \sqrt{1 - {e_1^2}}$. 

In the Hamiltonian, it is observed that $\Omega_2$ and $\varpi_1$ appear in the form of $\Omega_2^* = \Omega_2-\varpi_1$ (see Appendix \ref{sec:app}). For convenience, we introduce the following set of canonical variables:
\begin{equation}\label{Eq8}
\begin{aligned}
\theta_1 &= \Omega_2 - \varpi_1, \quad \Theta_1 = H_2,\\
\theta_2 &= \omega_2, \quad \Theta_2 = G_2,
\end{aligned}
\end{equation}
where $G_2$ and $H_2$ are Delaunay's action variables, defined by
\begin{equation*}
G_2= \sqrt{{\cal G} m_{12} a_2 (1-e_2^2)},\quad H_2 = G_2\cos{i_2},
\end{equation*}
with $m_{12} = m_1 + m_2$. As a result, the Hamiltonian governing the evolution of the inner and outer binaries can be written as
\begin{equation}\label{Eq9}
{\cal H} \left(\theta_1,\theta_2,\Theta_1,\Theta_2\right) = - {\dot \varpi _1} \Theta_1 + {\cal H}_{\rm sec} \left(\theta_1,\theta_2,\Theta_1,\Theta_2\right),
\end{equation}
which determines a two-degree-of-freedom (2-DOF) dynamical system. Accordingly, the equations of secular motion for the inner and outer binaries can be derived from the Hamiltonian canonical relations.

Furthermore, the GW dissipation drives the orbital decay of the inner BHB. This process is governed by the following coupled equations \citep{peters1964gravitational}:
\begin{equation}\label{Eq10}
\begin{aligned}
{\frac{{{\rm d}a_1}}{{{\rm d}t}}}  &=  - \frac{{64}}{5}\frac{{{{\cal G}^3}{m_1}{m_2}m_{12}}}{{{c^5}{a_1^3}{\eta_1^7}}}\left( {1 + \frac{{73}}{{24}}{e_1^2} + \frac{{37}}{{96}}{e_1^4}} \right),\\
{\frac{{{\rm d}e_1}}{{{\rm d}t}}}  &=  - \frac{{304}}{{15}} \frac{{{{\cal G}^3}{m_1}{m_2}m_{12}}}{{{c^5}{a_1^4}{\eta_1^5}}} e_1 \left( {1 + \frac{{121}}{{304}}{e_1^2}} \right).
\end{aligned}
\end{equation}
The corresponding analytical solution $a_1(e_1)$ is given by
\begin{equation}\label{Eq11}
    a_1(e_1) = {c_0} \frac{{{e_1^{\frac{12}{19}}}}}{{\eta_1^2}}{\left( {1 + \frac{{121}}{{304}}{e_1^2}} \right)^{\frac{870}{2299}}},
\end{equation}
where $c_0$ is an integration constant determined by the initial values of $a_1$ and $e_1$. 

In summary, the outer particle (representing a star) moves in a time-dependent (slowly varying) gravitational field produced by the inspiraling BHB. Specifically, the conservative dynamics are governed by the Hamiltonian (\ref{Eq9}), and the orbital decay of the inner BHB due to GW emission is described by Equation (\ref{Eq10}). In practice, this coupled system of differential equations is solved numerically using an adaptive Runge--Kutta--Fehlberg (RKF78) integrator \citep{fehlberg1968classical}. In the following sections, we investigate the dynamical behavior of the outer particles under three different configurations, including the quadrupole, coplanar octupole, and spatial octupole scenarios. Notably, all the hierarchical systems considered in this work satisfy the dynamical stability criterion \citep{holman1999long,mardling2001tidal}.

\begin{figure*}
\centering
\includegraphics[width=0.85\columnwidth]{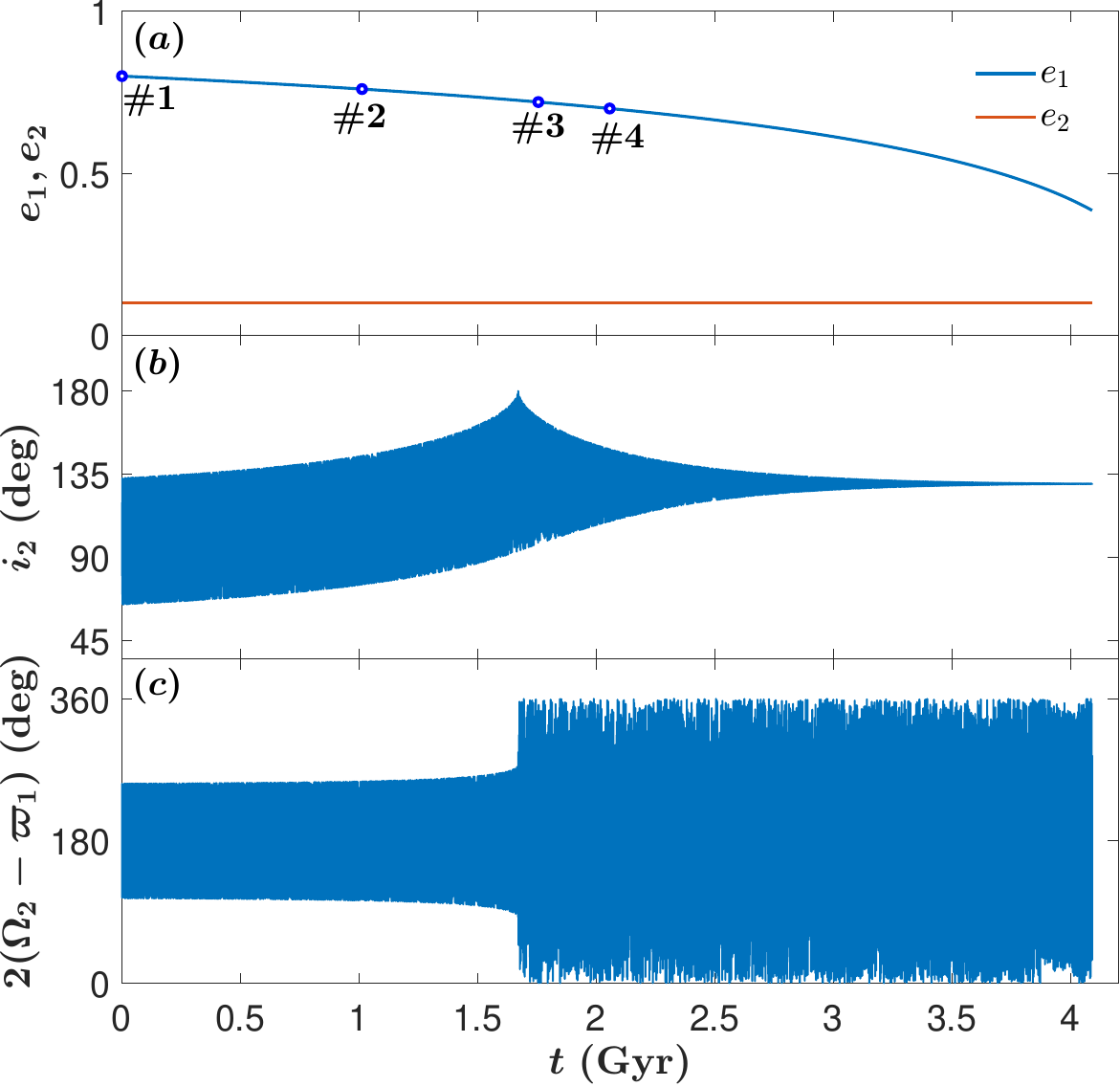}
\includegraphics[width=\columnwidth]{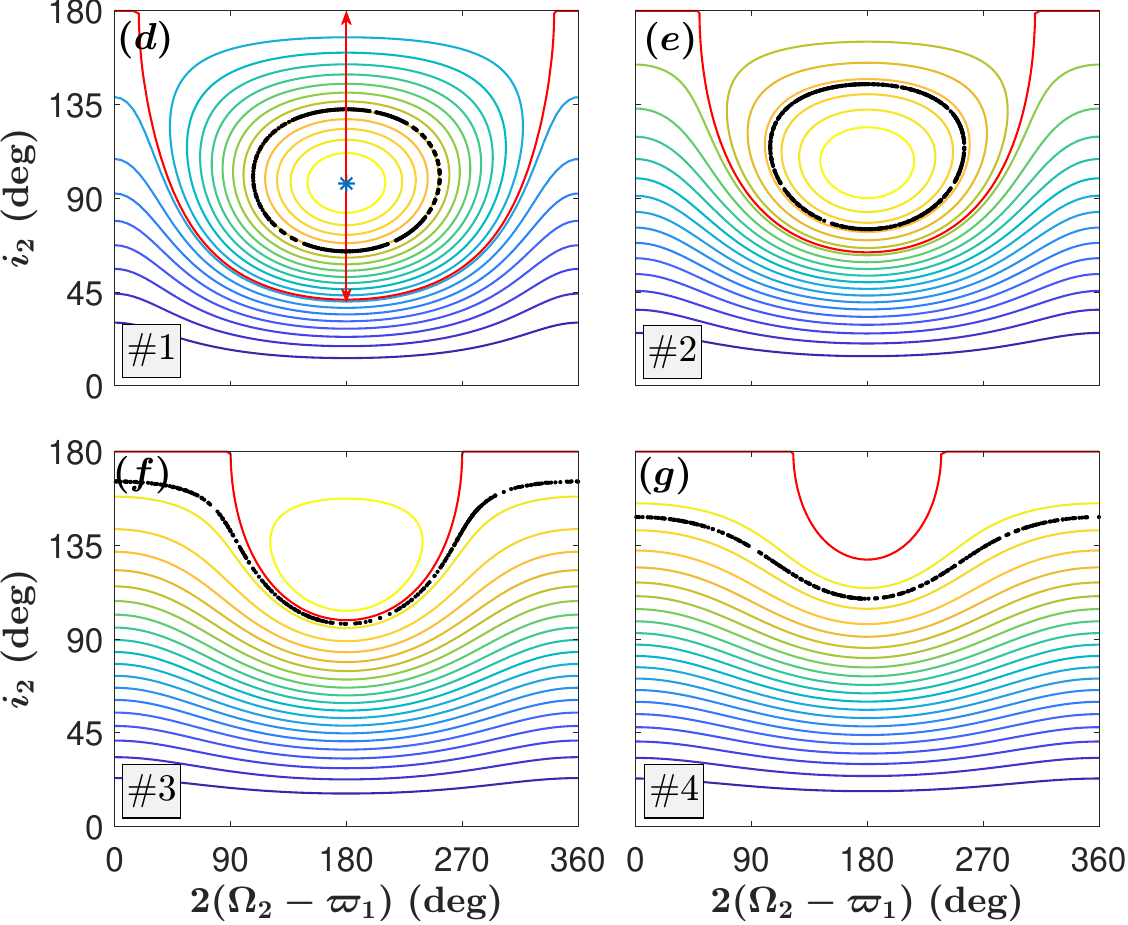}
\caption{Time histories of the eccentricties ($e_1$ and $e_2$), inclination $i_2$, and the argument $\sigma_1 = 2(\Omega_2 - \varpi_1)$ under the quadrupole-order model (panels a-c), and pseudo phase portraits at four different stages of the BHB's orbital decay marked by points from \#1 to \#4 (panels d-g). The inner BHB consists of equal-mass components \(m_1 = m_2 = 30\,M_{\odot}\), with initial state at \(a_{1,0}=0.4\,\mathrm{au}\) and \(e_{1,0}=0.8\). The outer particle starts from \(a_{2,0}=5\,\mathrm{au}\), \(e_{2,0}=0.1\), \(i_{2,0}=80^\circ\), and \(\sigma_{1,0}=120^\circ\). In panels (d-g), dynamical separatrices are shown in red lines, and the numerical trajectory considered in panels (a-c) is marked in black dots. The blue star in panel (d) stands for the center of quadrupole-order resonance, and the length of red arrow measures the resonant width. 
\label{fig:1}}
\end{figure*}

\begin{figure}
\includegraphics[width=0.85\columnwidth]{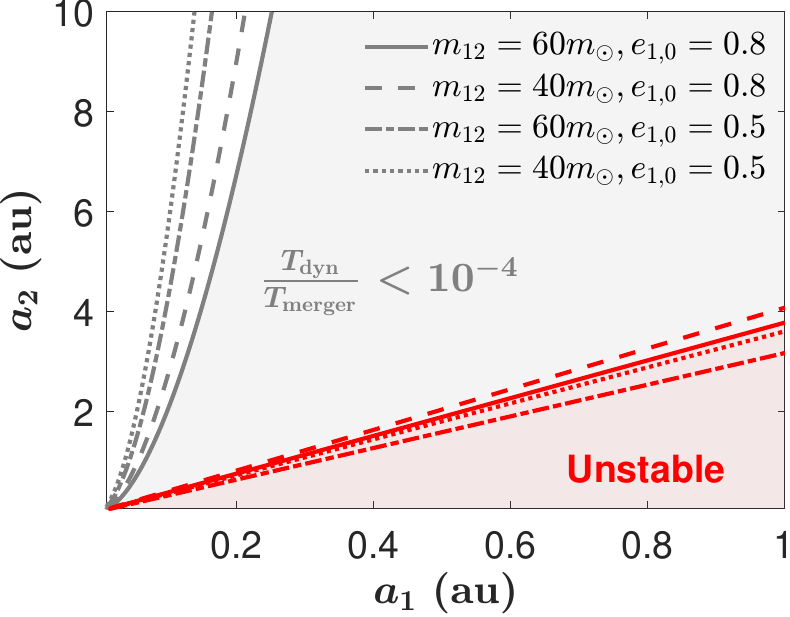}
\caption{The timescale ratio $T_{\rm dyn}/T_{\rm merger}$ in the $(a_1,a_2)$ plane. The gray contours mark where the ratio equals $10^{-4}$, while the red curves denote the instability boundaries, with different line styles corresponding to various mass parameters and initial eccentricities. The red-shaded region indicates the unstable zone. In the gray-shaded region where the ratio is below $10^{-4}$, systems farther from the $10^{-4}$ contour exhibit a stronger scale separation and thus better satisfy the adiabatic condition.}
\label{fig:new1}
\end{figure}

\section{Dynamics in the quadrupole model}
\label{sec:3}

In this section, we investigate the dynamical evolution of outer particles at the quadrupole level. It is a good approximation when the semimajor axis ratio is sufficiently small and/or the inner BHB is of equal mass.

Under the quadrupole model, a representative numerical example is presented in Figure \ref{fig:1}(a-c), where the caption provides a detailed description of system parameters and initial conditions. It is observed that, the eccentricity of the outer orbit, $e_2$, remains constant throughout the evolution, while the orbital inclination $i_2$ is gradually excited and eventually approaches a steady outcome (see Figure \ref{fig:1}(a\&b)). The conservation of $e_2$ can be understood as follows: at the quadrupole level, the Hamiltonian is independent of the argument of the pericenter $\omega_2$, and thus its conjugate momentum $G_2$ (and the outer eccentricity $e_2$) is conserved.

Figure \ref{fig:1}(c) shows that the argument $2\theta_1 = 2(\Omega_2-\varpi_1)$ exhibits libration during the stage of inclination growth, indicating that the outer particle is trapped within the quadrupole-order resonance. As the inner BHB undergoes orbital decay and circularization, the outer particle eventually transitions from libration to circulation, accompanied by the termination of inclination excitation.

To understand the quadrupole-order resonant dynamics, it is convenient to construct a Hamiltonian formulation in which the resonant angle serves as an independent angular coordinate. As a result, the quadrupole-order Hamiltonian can be written as
\begin{equation}\label{Eq12}
{\cal H} (\sigma_1, \Sigma_1, \Sigma_2) = - 2 {\dot \varpi _1} \Sigma_1 - {\cal C}_0 F_{\rm{quad}} (\sigma_1, \Sigma_1, \Sigma_2),
\end{equation}
which determines an integrable system with $(\sigma_1 = 2\theta_1,\Sigma_1 = \frac{1}{2} \Theta_1)$ as the unique pair of canonical variables, depending on the motion integral $\Sigma_2 (=\Theta_2)$. Remind that the variables $(\theta_1,\theta_2,\Theta_1,\Theta_2)$ are defined by Equation (\ref{Eq8}). 

Due to GW dissipation within the inner BHB, the semi-major axis $a_1$ and eccentricity $e_1$ gradually decrease. Usually, the orbital decay of the inner BHB occurs on a timescale much longer than that of the resonant oscillation, thus an adiabatic approximation is well justified. Specifically, the expressions for these two timescales are \citep{peters1964gravitational,lei2024dynamical}
\begin{equation}\label{Eqnew1}
\begin{aligned}
T_{\text{merger}}=&\frac{5a_{1,0}^4 c^5 \left(1-e_{1,0}^2\right)^{7/2}}{{256 {\cal G}^3m_1 m_2m_{12}^2}},\\
T_{\text{dyn}}= & \frac{16a_2^4 m_{12}^2}{a_1^2 m_1 m_2 {({\cal G}a_2 m_{12})^{1/2}}},
\end{aligned}
\end{equation}
where $a_{1,0}$ and $e_{1,0}$ are the initial orbital elements of the inner BHB. Figure \ref{fig:new1} shows the ratio of the two timescales across the $(a_1,a_2)$ plane, confirming a strong scale separation ($T_\text{merger} \gg T_\text{dyn}$) in the stable zone. Under this approximation, $a_1$ and $e_1$ can be treated as constant parameters at a given epoch, such that the entire evolution of test particle can be described by a sequence of quasi-static (or frozen) Hamiltonian systems \citep{henrard1982capture,neishtadt1984separation}. This framework enables us to analyze the evolution of the phase-space structure as the system parameters drift, and in particular, to track the growth and eventual disappearance of the libration island associated with the quadrupole-order resonance.

With the resonant Hamiltonian in hand, we are ready to visualize the resonant dynamics by constructing phase portraits. To illustrate the evolution of the test particle, four different stages of the BHB's orbital decay (marked by points from \#1 to \#4) are considered to produce phase-space structures, as shown in Figure \ref{fig:1}(d-g). In particular, the numerical trajectory considered in Figure \ref{fig:1}(a-c) is marked in black dots, showing good agreement with the level curves in the phase portraits. At each stage, the phase portrait represents a frozen snapshot of phase-space structures in the dissipative system. 

In general, as the test particle exits the resonant regime, the libration region shrinks and eventually vanishes, consistent with the transition from libration to circulation and thus the termination of inclination excitation. In the initial stage, the phase portrait exhibits a resonance island centered at $180^\circ$, and the test particle undergoes libration within the island, with the orbital inclination $i_2$ oscillating around the resonant center (see Figure \ref{fig:1}(d)). during the orbital decay of the inner BHB, the resonant region and the resonant center shift toward higher inclinations. During this migration, the particle is captured within the quadrupole-order resonance, and thus it gradually evolves towards a higher-inclination state. Particularly, when the eccentricity of the inner BHB decreases to $e_1 \approx 0.72$ (see Figure \ref{fig:1}(f)), the particle escapes from the resonance, and since then the inclination gradually approaches a constant value (see Figure \ref{fig:1}(b)).

The resonant Hamiltonian (\ref{Eq12}) has an explicit expression, enabling us to readily identify the resonant center and the resonance boundary, given by
\begin{equation}\label{Eq13}
\begin{aligned}
\cos{i_{{\rm{cen}}}} =& { - \frac{{{\cal G}{m_{12}^3}}}{{{m_1}{m_2}}}\frac{{4a_2^{7/2}{\eta_2^4}}}{{a_1^{9/2}{c^2}\eta_1^2\left( {1 + 4e_1^2} \right)}}},\\
\cos{i_{{\rm{sep}}}} = & { - \frac{{4a_2^3{\eta_2^{3}}m_{12}}}{{a_1^{9/2}{c^2}\eta_1^2(1 + 4e_1^2){m_1}{m_2}}}}  \\
&\times \left( {{\cal G}{m_{12}^2} {{a_2^{1/2}}\eta_2} } \right. + a_1^{5/2}{c^2} {\eta_1^2}{\left.  \sqrt{{\cal A} - {\cal B}} \right)},
\end{aligned}
\end{equation}
where
\begin{equation*}
{\cal A} =\frac{{5a_1^4e_1^2(1 + 4e_1^2)m_1^2m_2^2}}{{16a_2^6{{\eta_2^6}}{m_{12}^2}}},\quad {\cal B}=\frac{5{{\cal G}^2}{m_{12}^4}{{a_2}e_1^2\eta_2^2}}{{a_1^5{c^4}{\eta_1^6}}}.
\end{equation*}

\begin{figure*}
\centering
\rotatebox{90}{\resizebox{0.428\textwidth}{!}{\plotone{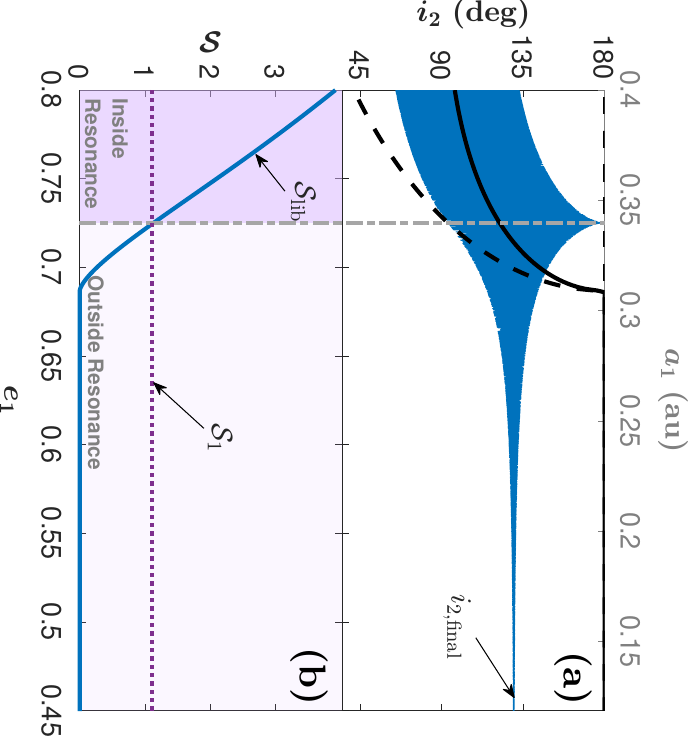}}}
\includegraphics[width=\columnwidth]{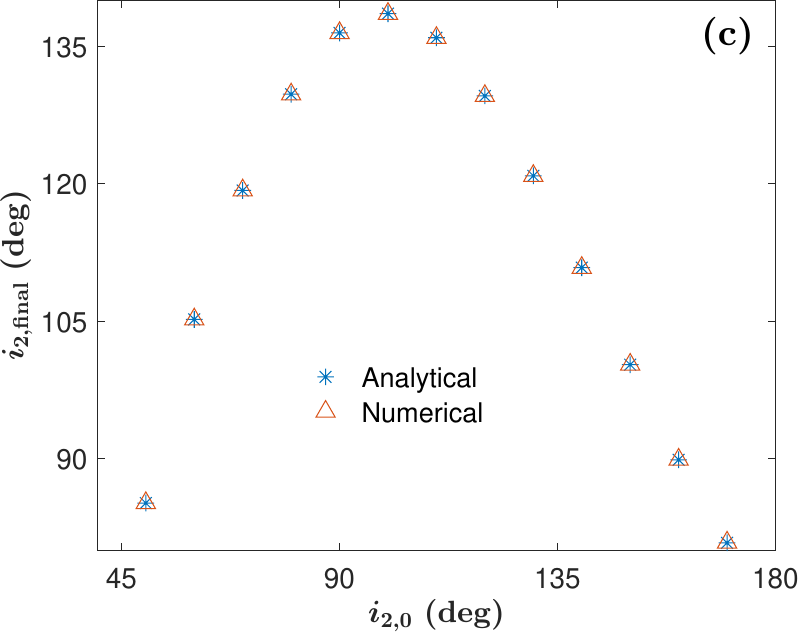}
\caption{In panel (a), analytical results derived from the quadrupole-order Hamiltonian framework, including the center and boundaries of quadrupole-order resonance given by Equation (\ref{Eq13}), are shown in the spaces of $(e_1,i_2)$ and $(a_1,i_2)$ by black solid and dashed lines, respectievly. For convenience of comparision, the numerical trajectory considered in Figure \ref{fig:1} is plotted here by blue curve, with the steady outcome of inclination marked by $i_{2,\rm final}$. Panel (b) shows the phase-space area of the libration region ${\cal S}_\text{lib}$ and that of the considered example ${\cal S}_1$ as functions of \(e_1\) and/or \(a_1\). The grey vertical dashed line denotes the critical moment when ${\cal S}_1 = {\cal S}_\text{lib}$. Such a critical line divides the evolution into two distinct phases: resonance capture stage (inside resonance), and resonance escape stage (outside resonance). Panel (c) illustrates the comparision of analytical and numerical relations between $i_{2,\rm final}$ and $i_{2,0}$. Please refer to Equation (\ref{Eq15}) for analytical expression of $i_{2,\rm final}$.
\label{fig:2}}
\end{figure*}

Figure \ref{fig:2}(a) shows the analytical results of resonant center and boundaries, together with the numerical trajectory in the $(e_1, i_2)$ plane. We find that the numerical trajectory oscillates around the characteristic curve of resonant center, and the particle's inclination is adiabatically pumped to a high level. This agreement confirms that the inclination excitation is due to the quadrupole-order resonance capture. 

To understand the eventual breakdown of the resonance, we invoke the theory of adiabatic invariance \citep{henrard1982capture,neishtadt1984separation}. As discussed above, the orbital decay of the inner BHB occurs on a timescale much longer than that of the resonant oscillation, allowing the dynamics to be described by a sequence of quasi-static Hamiltonians. In such systems, the phase-space area enclosed by the trajectory \citep{henrard1982capture}
\begin{equation}\label{Eq14}
    {\mathcal S}_1 = \oint \cos i_2 \, {\rm d}\sigma_1
\end{equation}
is approximately conserved, acting as an adiabatic invariant. Here $\sigma_1 = 2(\Omega_2 - \varpi_1)$ is the argument of the quadrupole-order resonance.

During the orbital decay of the inner BHB, the structure of Hamiltonian changes continuously, leading to a gradual shrinkage of the libration region in phase space. While the test particle remains in resonance, its trajectory is confined within this region and its phase-space area remains approximately constant. However, once the area of the resonance island becomes smaller than that enclosed by the test particle trajectory, the resonance can no longer confine the motion. As indicated by the gray dashed line, resonance escape occurs when the trajectory touches the resonant boundary (at the moment marked by the grey dashed line).

In Figure \ref{fig:2}(b), we plot the evolution of the area of the resonant region, denoted by ${\cal S}_\text{lib}$ (phase-space area bounded by the dynamical separatrix). It is evident that this area continuously decreases with the orbital decay of the inner BHB. Therefore, for a test particle initially located inside the resonance island, it inevitably escapes from resonance at some point during the evolution. Conversely, for a test particle initially outside the resonance island, its phase-space area is already larger than the resonant region from the outset, and it will never be captured into the resonance island to experience inclination excitation. 

In Figure \ref{fig:2}(a-b), the grey vertical dashed line denotes the critical moment at which it holds ${\cal S}_1 = {\cal S}_\text{lib}$ (touching the dynamical separatrix of quadrupole-order resonance). This critical line separates the entire evolution into two distinct phases: resonance capture stage (inside resonance), and resonance escape stage (outside resonance).

After escaping from the resonance, the inclination gradually approaches a constant value. Under the adiabitic approximation, this final inclination can be estimated by conserving ${\cal S}_1$ as follows:
\begin{equation}
    \cos\left({i_{\text{2,final}}}\right)=\frac{{\cal S}_1}{2\pi}-1
    \label{Eq15}
\end{equation}
where ${\cal S}_1$ is the phase-space area determined by the initial condition. Thus, the final inclination $i_{2,\rm final}$ is determined by its initial value $i_{2,0}$. Figure \ref{fig:2}(c) presents a comparison between the final inclinations obtained from numerical integration and those predicted by the analytical solution (\ref{Eq15}) as a function of the initial inclination $i_{2,0}$. It is observed that the analytical results agree excellently with the numerical ones.

It should be noted that, within the adiabatic theory, when the trajectory crosses a separatrix, the adiabatic invariant is no longer strictly conserved. Its relative change satisfies $\Delta {\cal S}_1/{\cal S}_1 \sim \epsilon |\ln\epsilon|$, where $\epsilon \equiv T_{\rm dyn}/T_{\rm merger}$ \citep{tennyson1986change}. For the parameters adopted in this work, $\epsilon$ is sufficiently small such that the deviation induced by separatrix crossing is negligible. In fact, for the transition from Figure \ref{fig:1}(e) to Figure \ref{fig:1}(f), there is a relative change of only $0.14\%$ in the adiabatic invariant. Therefore, although Equation (\ref{Eq15}) is derived under approximation, it remains accurate enough for the present purpose.

\begin{figure*}
\centering
\includegraphics[width=\columnwidth]{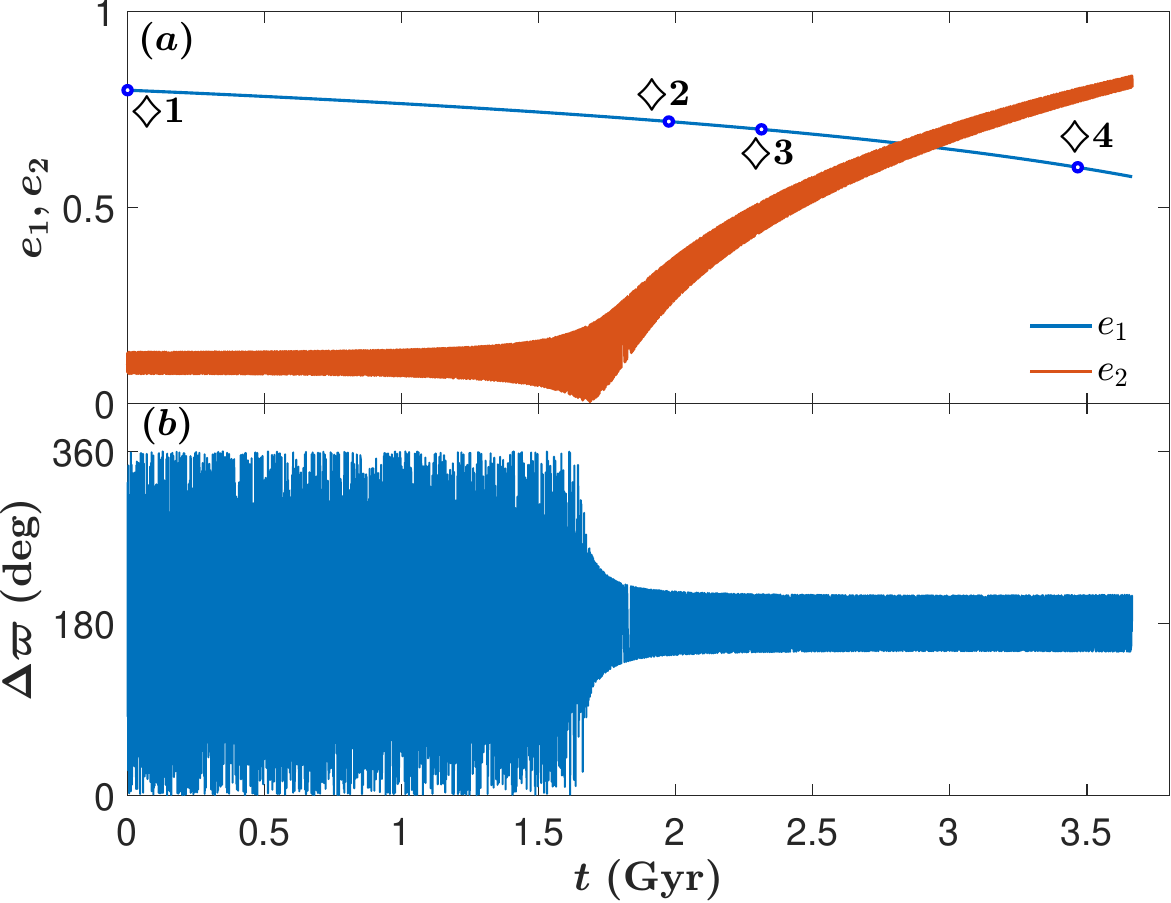}
\includegraphics[width=0.9\columnwidth]{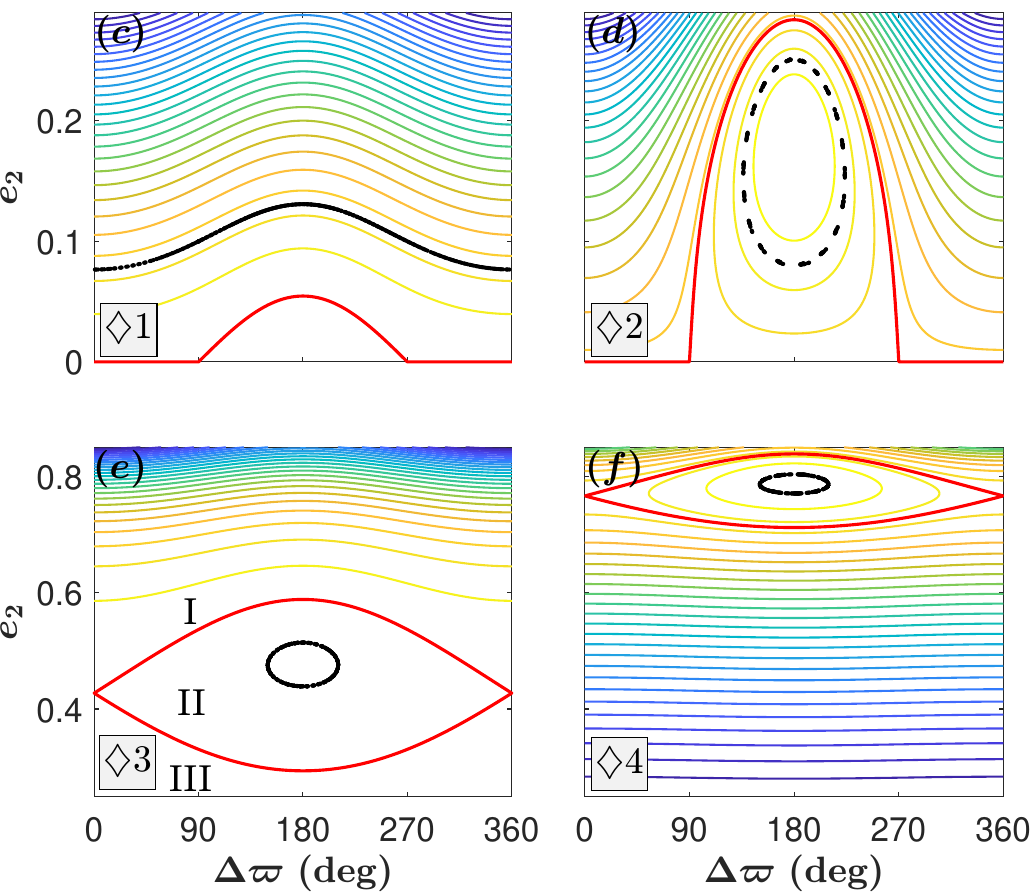}
\caption{Time histories of eccentricities ($e_1$ and $e_2$) and argument of $\Delta\varpi=\varpi_2 - \varpi_1$ under the coplanar octupole-order limit (panels a-b), and pseudo phase portraits at four different stages of the BHB's orbital decay marked by points from $\diamondsuit\,1$ to $\diamondsuit\,4$ (panels c-f). The BHB holds masses \(m_1 = 20\,M_{\odot}\) and \(m_2 = 40\,M_{\odot}\), with initial state at \(a_{1,0} = 0.4\,\mathrm{au}\) and \(e_{1,0} = 0.8\). The outer particle (star) starts from \(a_{2,0} = 5\,\mathrm{au}\), \(e_{2,0} = 0.1\), and \(\Delta\varpi_0 = 90^\circ\). In panels (c-f), the numerical trajectory considered in panels (a-b) is marked by black dots. Dynamical separatrix (red line) divides phase space into libration regime (zone II) and circulation regimes (zones I and III).
\label{fig:3}}
\end{figure*}

\section{Dynamics in the coplanar octupole model}
\label{sec:4}

In this section, we discuss the dynamics in the coplanar octupole-order configuration. Figure \ref{fig:3}(a\&b) shows a representative example of orbital evolution in coplanar triples. A similar behaviour of eccentricity excitation was observed in coplanar triples by \cite{liu2024extreme}.

From Figure \ref{fig:3}(a), we can see that the eccentricity of the outer orbit, \(e_2\), undergoes significant growth. In particular, during the eccentricity-excitation stage, the argument
\begin{equation*}
    \Delta \varpi = \varpi_2 - \varpi_1
\end{equation*}
exhibits librating behavior, as shown in Figure \ref{fig:3}(b).

Similar to the quadrupole-order resonance discussed in Section \ref{sec:3}, we still construct a Hamiltonian formulation in which the resonant angle explicitly appears as an independent variable. To this end, we introduce \(\Delta \varpi = \varpi_2 - \varpi_1\) as a canonical coordinate through the following canonical transformation:
\begin{equation}\label{Eq16}
    \begin{aligned}
        &\sigma_1 = \theta_1 + \theta_2 = \Delta \varpi, \quad \Sigma_1 = \Theta_1 = H_2,\\
        &\sigma_2 = \theta_2 = \omega_2,\quad \Sigma_2 = \Theta_2 - \Theta_1 = G_2 - H_2.
    \end{aligned}
\end{equation}
It is noted that, in the coplanar configuration ($i_2 = 0$), it holds $G_2 = H_2$. This means $\Sigma_2 = 0$. Under the new set of canonical variables, the octupole-order Hamiltonian in the coplanar configuration becomes
\begin{equation}\label{Eq17}
{\cal H} (\sigma_1,\Sigma_1) = - {\dot \varpi _1} \Sigma_1 + {\cal H}_{\rm sec} (\sigma_1,\Sigma_1),
\end{equation}
which determines an integrable system. An equivalent Hamiltonian formulation can be found in \cite{liu2024extreme}. 

Under the adiabatic approximation, we construct phase-space portraits in the \((\Delta \varpi, e_2)\) plane at four different stages of the BHB's orbital decay (marked by points from $\diamondsuit\,1$ to $\diamondsuit\,4$), as shown in Figure \ref{fig:3}(c-f).  In each portrait, the colored curves are level curves of the Hamiltonian, and the red curve represents the dynamical separatrix. Unlike the quadrupole case, the phase space is divided into two/three distinct regions, denoted by I, II (and III) from the top to bottom, as shown in Figure \ref{fig:3}(e).

Thanks to the phase portraits, the mechanism of eccentricity excitation can now be understood as follows. As the inner BHB's orbit decays, the resonance region gradually shifts toward higher outer eccentricities. When the particle enters this expanding resonance region, it can be captured into the apsidal precession resonance. Once captured, the outer eccentricity \(e_2\) is adiabatically pumped to a high value. This process continues until the inner BHB merges or the resonance region eventually shrinks.

\begin{figure*}
\hspace*{\fill}
\begin{minipage}[b]{0.398\textwidth}
\centering
\rotatebox{90}{\resizebox{\linewidth}{!}{\plotone{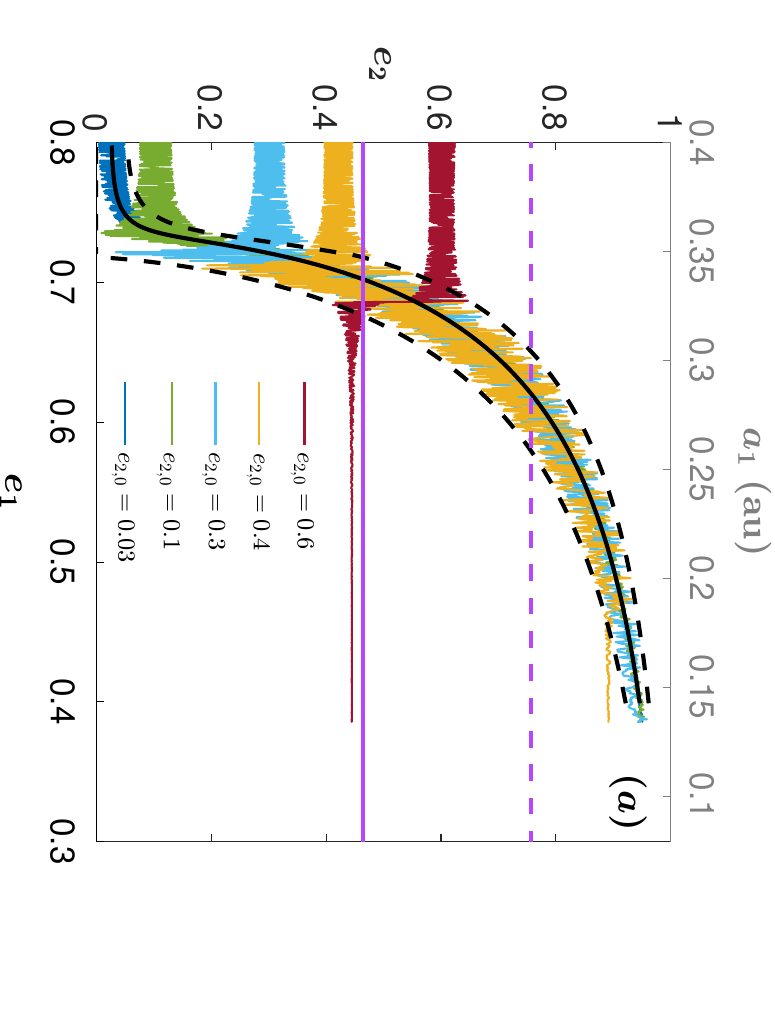}}}
\end{minipage}
\hspace{5pt}
\begin{minipage}[b]{0.478\textwidth}
\centering
\plotone{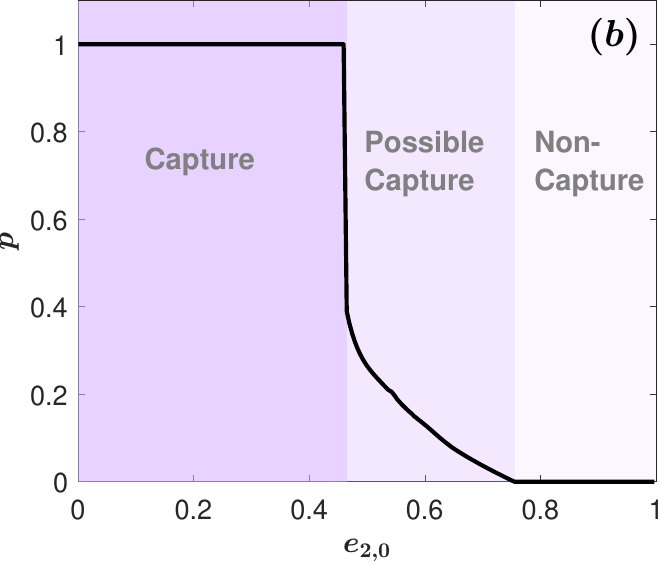}
\end{minipage}
\hspace*{\fill}
\caption{In panel (a), analytical results derived from the coplanar octupole Hamiltonian given by Equation (\ref{Eq17}), including the center and boundaries of apsidal precession resonance, are shown in the space of $(e_1,e_2)$ by black solid and dashed lines, respectievly. For convenience of comparision, numerical trajectories starting from different eccentricities are plotted as background. Panel (b) displays the probability of resonance capture $p$ as a function of \(e_{2,0}\). Accordingly, the space with $p=1$ is termed as capture, the space with $p=0$ as non-capture, and the remaining space as possible capture. The boundaries of capture, possible capture and non-capture regimes are marked by the purple (solid and dashed) lines in panel (a).  
\label{fig:4}}
\end{figure*}

Similar to the quadrupole case discussed in Section \ref{sec:3}, we predict the evolutionary trend of the test particle by computing the resonant center and the separatrix. To validate our theoretical framework, we numerically solve the full secular equations for different initial outer eccentricities \(e_{2,0}\), as shown in Figure \ref{fig:4}(a). For each case, we compare the trajectory in the \((e_1, e_2)\) plane with the resonant center and boundaries derived from the Hamiltonian. We find that all numerical solutions undergo eccentricity excitation precisely along the migrating curve of the resonant center, marked by the black solid line.

The capture and/or non-capture behaviors observed in Figure \ref{fig:4}(a) can be understood by resonance capture probability, defined by \citep{henrard1982capture,henrard1993adiabatic} \begin{equation}\label{Eq19}
    p = P_{\text{I} \to \text{II}}  = -\frac{\partial S_\text{II}}{\partial S_\text{I}},
\end{equation}
where $S_{\text{I}}$ and $S_{\text{II}}$ are the phase-space areas of region I and II (see Figure \ref{fig:3}(e)). Figure \ref{fig:4}(b) presents the probability of resonance capture as a function of the initial outer eccentricity $e_{2,0}$. It shows that the parameter space can be divided into three distinct regimes from the left to right: (i) a \emph{capture} regime, where the system is always captured into the libration zone and experiences strong eccentricity growth; (ii) a \emph{possible capture} regime, where the capture probability is between 0 and 1, possibly depending on the initial phase-space trajectory (e.g., the initial $\Delta\varpi_0$); and (iii) a \emph{non-capture} regime, where the system always passes through the lower separatrix and fails to be captured, resulting in no eccentricity excitation. The boundaries of these distinct regimes are marked by purple (solid and dashed) lines in Figure \ref{fig:4}(a). 

\begin{figure}
\centering
\includegraphics[width=\columnwidth]{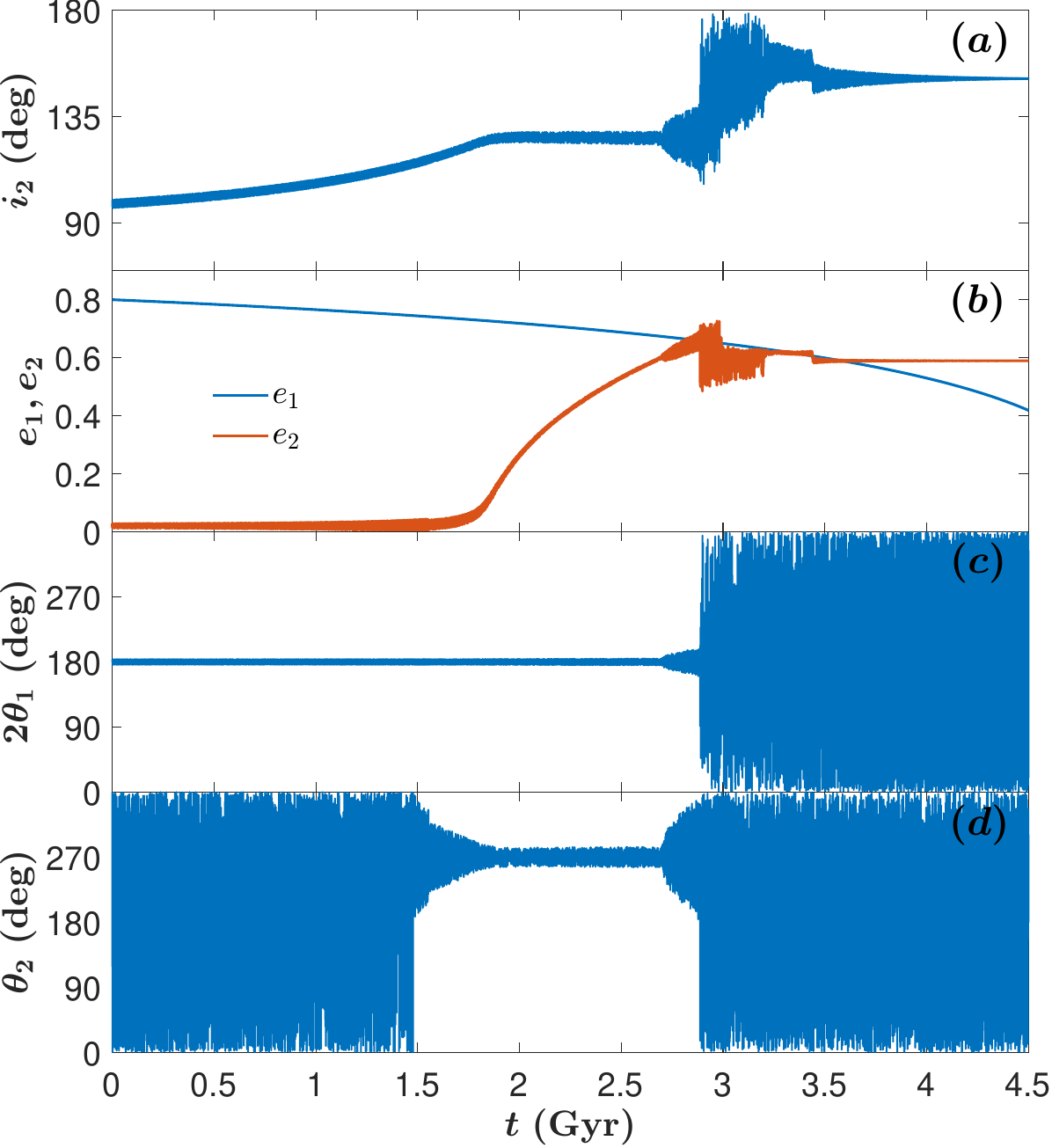}
\caption{Time histories of the inclination $i_2$, eccentricities ($e_1$ and $e_2$), and arguments of $2\theta_1 = 2(\Omega_2-\varpi_1)$ and $\theta_2 = \omega_2$ for an outer particle around merging BHB in the spatial octupole-order configuration. The inner BHB has masses \(m_1 = 20\,M_{\odot}\) and \(m_2 = 40\,M_{\odot}\), with initial state at \(a_{1,0}=0.4\,\mathrm{au}\), and \(e_{1,0}=0.8\). The outer particle starts with \(a_{2,0}=5\,\mathrm{au}\), \(e_{2,0}=0.02\), \(i_{2,0}=100^\circ\), \(\theta_{1,0}=90^\circ\), and \(\theta_{2,0}=0\).
\label{fig:5}}
\end{figure}

\section{Dynamics in the spatial octupole model}
\label{sec:5}

In this section, we extend the special analysis made in the previous two sections to the spatial octupole configuration, which exhibits rich dynamical behavior due to the presence of both eccentricity and inclination degrees of freedom.  

Figure \ref{fig:5} presents a representative numerical example. It is observed that the inclination $i_2$ is first excited from its initial state while the eccentricity $e_2$ remains nearly circular. After the inclination excitation saturates, the system enters a second stage, during which the eccentricity begins to grow, while the inclination remains approximately constant. Time histories of the angular coordinates $2\theta_1$ and $\theta_2$ are shown in panels (c) and (d), respectively. We can see that during the first (inclination-excitation) stage, the argument $2\theta_1$ librates around $180^{\circ}$, and during the second (eccentricity-excitation) stage both the arguments $2\theta_1$ and $\theta_2$ librate.

To understand this behavior observed in Figure \ref{fig:5} analytically, we calculate the equilibrium states under the secular Hamiltonian at the octupole level, where the Hamiltonian is given by Equation (\ref{Eq2}). Stable equilibrium states are equivalent to resonant centers, determined by the stationary conditions of the Hamiltonian,
\begin{equation}\label{Eq20}
    \frac{\partial \mathcal{H}}{\partial e_2} = 0, \qquad
    \frac{\partial \mathcal{H}}{\partial i_2} = 0,
\end{equation}
where the angular variables are fixed at $2\theta_1 = \pi$ and $\theta_2 = 3\pi/2$, corresponding to their respective resonant centers. 

\begin{figure*} 
\hspace*{\fill}
\begin{minipage}[b]{0.51\textwidth}
\centering
\plotone{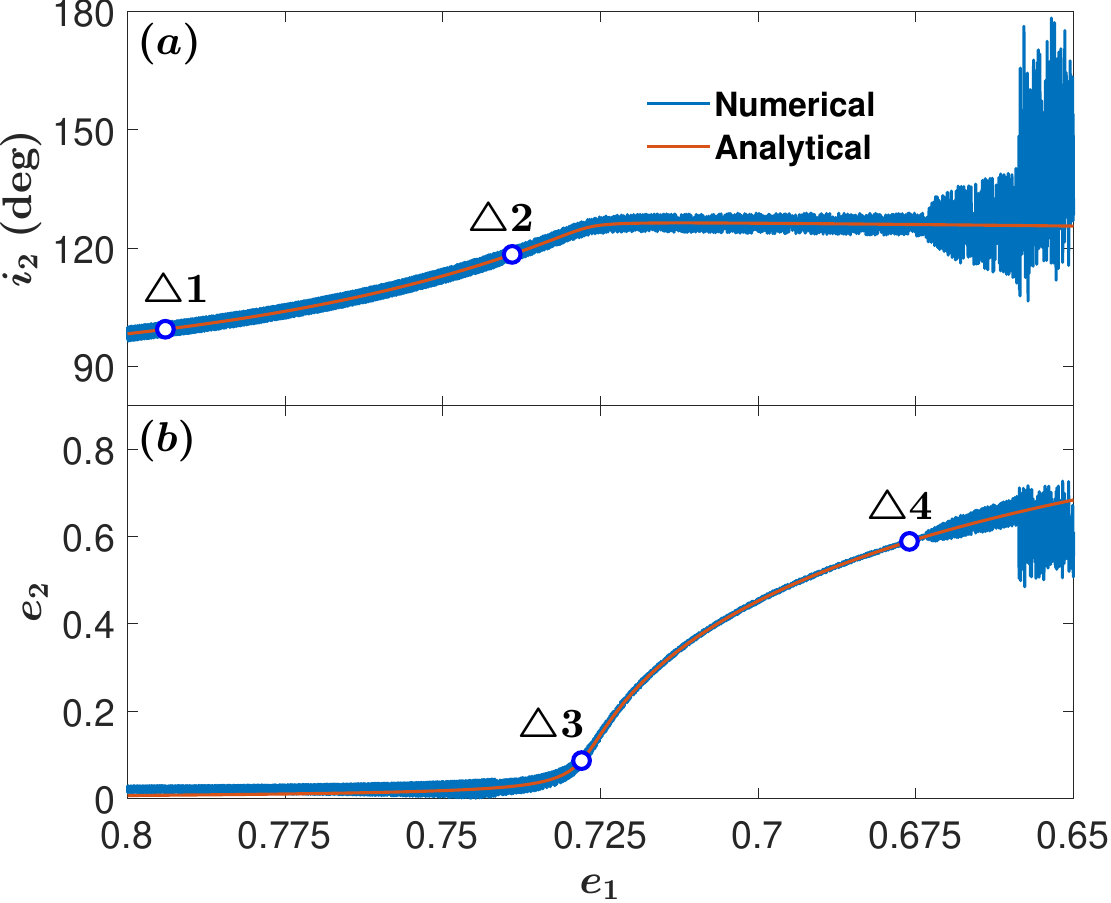}
\end{minipage}
\hspace{-25pt}
\begin{minipage}[b]{0.51\textwidth}
\centering
\plotone{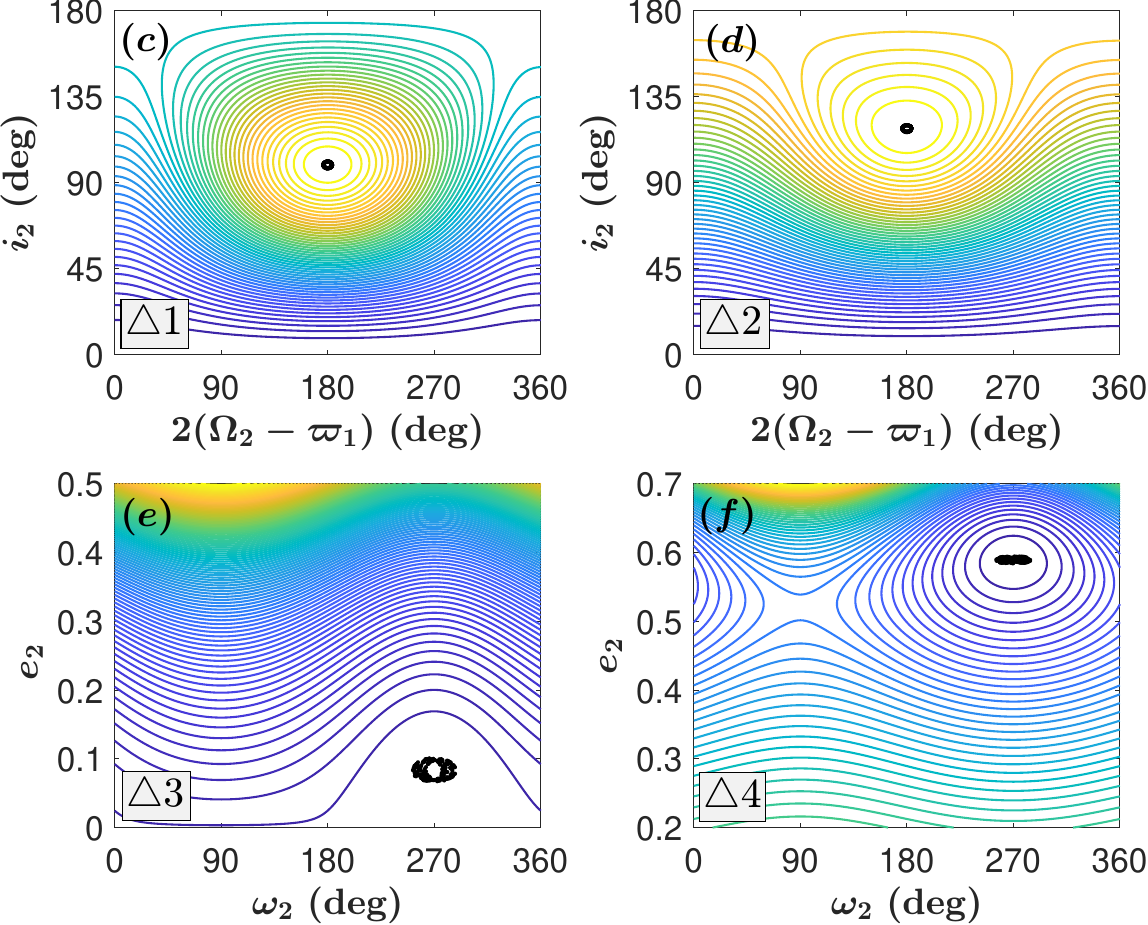}
\end{minipage}
\hspace*{\fill}
\caption{The distribution of (stable) equilibrium points and a numerical trajectory (see panels a-b), and dynamical structures (or representative planes) at four different stages of the BHB's orbital decay marked by points from $\triangle\,1$ to $\triangle\,4$ (see panels c-f). The numerical trajectory is the same as that in Figure \ref{fig:5}, and it is marked in black dots in panels (c-f).
\label{fig:6}}
\end{figure*}

By solving Equation (\ref{Eq20}), we obtain the distribution of (stable) equilibrium points of 2-DOF Hamiltonian model as the inner BHB's orbit decays. In particular, the inclination $i_2$ and eccentricity $e_2$ of equilibrium points are shown in Figure \ref{fig:6}(a\&b) as functions of the inner orbital eccentricity $e_1$ (see the red lines). For convenience, the numerical trajectory considered in Figure \ref{fig:5} is plotted in blue lines as background. We can see that, the numerical trajectory closely follows the characteristic curves of (stable) equilibrium solutions in the $(e_1,i_2)$ and $(e_1,e_2)$ planes. 

The agreement observed in Figure \ref{fig:6}(a-b) indicates that the numerical evolution under the general configuration is an adiabatic process. To illustrate the resonance-induced mechanism, we plot the phase‑space diagrams (or representative planes) of the 2-DOF Hamiltonian model in Figure \ref{fig:6}(c-f), corresponding to four different stages of the BHB's orbital decay marked by points from $\triangle\,1$ to $\triangle\,4$. In particular, the points $\triangle\,1$ and $\triangle\,2$ are located in the inclination-excitation stage, while the points $\triangle\,3$ and $\triangle\,4$ are located in the eccentricity-excitation stage.

\begin{figure}
\centering
\includegraphics[width=\columnwidth]{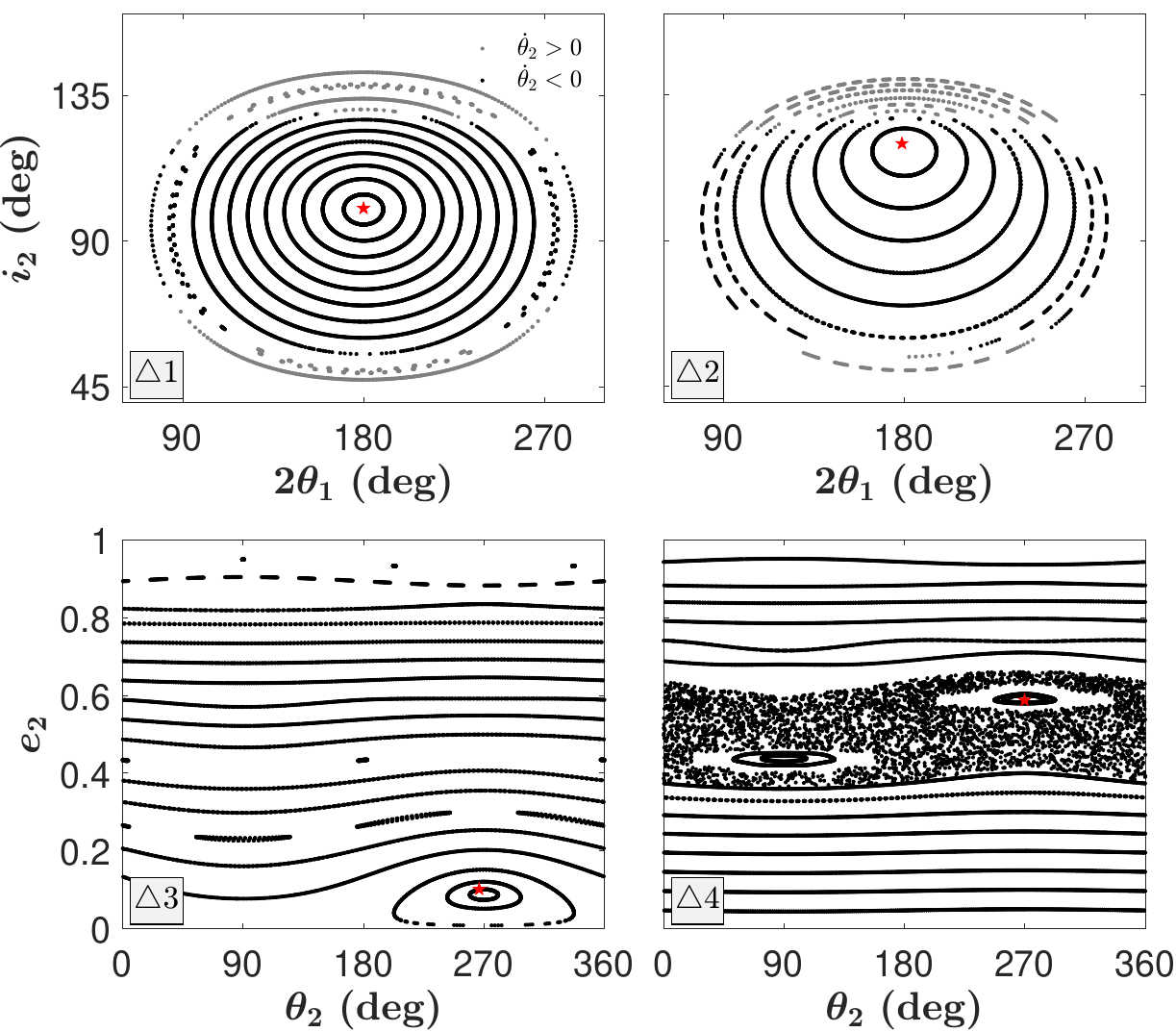}
\caption{Poincar\'e sections corresponding to the four representative stages marked in Figure \ref{fig:6}. The red star marks the position of the system on the Poincar\'e section. Different colors denote trajectories sampled in different directions across the section.}
\label{fig:new2}
\end{figure}

In Figure \ref{fig:new2}, we construct Poincaré sections corresponding to the four representative stages marked in Figure \ref{fig:6}. The Poincar\'e sections are defined respectively by $\theta_1=\pi/2, \dot\theta_1<0$ and $\theta_2=\pi/2$. The red pentagram in each section marks the system's location at the corresponding stage. As seen in Figure \ref{fig:new2}, the dynamical structures appearing in Figure \ref{fig:6} are in good agreement with the Poincaré sections. During the inclination excitation phase, both $\triangle\,1$ and $\triangle\,2$ exhibit a resonance island centered at $2\theta_1=180^\circ$. During the eccentricity excitation phase, $\triangle\,3$ and $\triangle\,4$ have a resonance island centered at $\theta_2=270^\circ$; however, at the end of the evolution, chaotic regions appear around the resonance island in $\triangle\,4$.

These results show that the complex evolution of the system can be well understood as a sequence of resonance-driven phases. The system is first captured into a quadrupole-order resonance (inclination-type resonance), leading to inclination excitation (see Figure \ref{fig:6}(c\&d)). After being subsequently captured into the inverse ZLK resonance (eccentricity-type resonance), both resonances operate simultaneously. However, during this stage, the inclination remains approximately constant, while the eccentricity continues to be excited under the influence of the inverse ZLK resonance (see Figure \ref{fig:6}(e\&f)). This sequential yet coexisting resonance structure provides a coherent dynamical framework for understanding the coupled evolution at the octupole level in an inclined configuration.

\section{Conclusions and Discussion} \label{sec:6}

In this work, we have systematically explored the secular dynamical evolution of a stellar test particle orbiting a merging BHB, focusing on the resonance-driven excitation of eccentricity and/or inclination. By coupling the dissipative inspiral of the inner BHB with secular perturbations up to the octupole order, we have identified distinct resonant capture channels, offering a unified dynamical picture. Our primary results are summarized as follows:
\begin{itemize}
\item \textbf{Inclination excitation.} At the quadrupole level, the inclination of the outer orbit grows through adiabatic capture into the quadrupole-order resonance with the resonant argument $2(\Omega_2 - \varpi_1)$. As the inner BHB undergoes orbital decay, the resonant island migrates toward a higher inclination, dragging the test particle along until the phase-space area can no longer sustain the adiabatic invariant. This process provides a clear picture of resonant capture and separatrix crossing in a slowly varying (integrable) Hamiltonian system.
\item \textbf{Eccentricity excitation.} When the octupole-order correction is included, the coplanar configuration exhibits significant eccentricity excitation driven by the apsidal precession resonance with the argument $\Delta\varpi = \varpi_2 - \varpi_1$. The libration zone expands toward higher $e_2$ during the circularization of the inner orbit, leading to a probabilistic capture scenario. We have quantified the capture probability as a function of the initial eccentricity $e_{2,0}$, delineating capture, possible capture, and non-capture regimes. This analysis clarifies the conditions under which extreme eccentricity growth can occur in coplanar hierarchical triples.
\item \textbf{Coupled excitation of eccentricity and inclination.} The general case—the spatial octupole configuration—reveals a two-stage evolutionary pathway. Initially, the inclination is excited by the quadrupole-order resonance while $e_2$ remains nearly constant. Once the inclination stabilizes, the system enters the inverse ZLK resonance, yielding significant eccentricity excitation. This sequential behavior unifies the previously separate mechanisms of inclination and eccentricity excitation into a single coherent framework driven by the slow inspiral of the inner BHB.
\end{itemize}

The resonance-driven excitation of stellar eccentricity and/or inclination uncovered in this work provides a natural dynamical tracer for hidden merging BHBs. A star that has undergone secular resonance capture will carry a distinctive orbital signature—an anomalously high eccentricity and/or a misaligned orbit—that cannot be easily explained by isolated stellar evolution or binary interactions alone. Such stars, if identified in astrometric or spectroscopic surveys, could serve as indirect signposts pointing to the presence of an unseen inspiraling compact binary in their vicinity. This offers an alternative pathway to locate GW progenitors prior to merger.

Beyond the excitation of individual stellar eccentricities and inclinations, the mechanism may leave observable imprints in the surrounding stellar population. Stars affected by this process may exhibit dynamically unusual orbital properties, such as an excess of highly eccentric, highly inclined, or retrograde orbits relative to the binary orbital plane. Moreover, if this process operates on a population of stars, it may lead to statistical signatures in their orbital distributions, such signatures may provide additional constraints on the dynamical history of compact binaries and their surrounding environments.

Furthermore, the mechanism only requires a slowly evolving inner binary, regardless of the specific dissipation channel. Thus, similar resonance excitation may operate in hierarchical stellar triples and circumbinary planetary systems, where secular resonances can drive eccentricity and inclination excitation through tidal or other forms of dissipation \citep{farhat2025capture,liu2026planetary}. It offers additional insights into the origin of dynamically excited stellar companions and misaligned or eccentric planetary systems, complementing existing scenarios such as dynamical scattering, Kozai–Lidov oscillations, and disk-driven evolution.

It should be noted that the dynamical model considered in this work is formulated in the restricted limit. When the outer body has a finite mass, its gravitational influence upon the inner BHB cannot be ignored. In this case, the eccentricity decay rate of the inner BHB may be reduced due to the angular momentum exchange. However, this back-reaction does not qualitatively alter the excitation of eccentricity and inclination of the outer orbit \citep{liu2024extreme}. Furthermore, when the outer orbit becomes extremely eccentric, the outer pericenter passage time may become comparable to the inner orbital period, and thus the standard double-averaging approximation may be challenged. Within such quasi-hierarchical regimes, second-order perturbation techniques may be required \citep{hamers2019analytic,hamers2019analyticb,ginat2026dynamical}.


\begin{acknowledgments}
We thank Dong Lai, Bin Liu, and Xiyun Hou for helpful discussions and suggestions. This work is financially supported by the National Natural Science Foundation of China (Nos. 12573063, and 12233003).
\end{acknowledgments}

 \appendix
 \section{Quadrupole and Octupole Terms}
 \label{sec:app}
 
The quadrupole- and octupole-order terms in Equation (\ref{Eq2}) are
\begin{equation}\label{EqA1}
\begin{aligned}
F_{\rm{quad}} = &\frac{1}{\eta_2^3}\left\{ {(2 + 3e_1^2)(3\theta^2-1)}{ + 15 e_1^2 (1 - {\theta^2})\cos (2{\Omega _2^*})} \right\},
\end{aligned}
\end{equation}
and
\begin{equation}\label{EqA2}
\begin{aligned}
F_{\rm{oct}}=&\frac{{15{e_1}{e_2}}}{{32{\eta_2^5}}}\left\{(4 + 3e_1^2) \right.
 [(1 - 11\theta - 5\theta^2 + 15\theta^3)\cos ({\Omega _2^*} - {\omega _2})+(1 + 11\theta - 5\theta^2 - 15\theta^3)\cos ({\Omega _2^*} + {\omega_2})]\\
&{-35e_1^2[(1 - \theta - \theta^2 + \theta^3)\cos (3{\Omega _2^*} - {\omega _2})}\left.{+{(1 + \theta - \theta^2 - \theta^3)}\cos (3{\Omega _2^*} + {\omega _2})]} \right\}
\end{aligned}
\end{equation}
where ${\Omega _2^*} = {\Omega_2} - {\varpi_1}$, $\theta=\cos{i_2}$, and $\eta_2 = \sqrt{1-e_2^2}$.

\bibliography{reference}{}
\bibliographystyle{aasjournalv7}



\end{document}